\documentstyle[preprint,aps,epsf]{revtex}
\tighten

\begin{document}
\draft
\title{Supersymmetry approach to the random heteropolymer theory}
\author{Alexander I. Olemskoi}
\address{Physical Electronics Department, Sumy State University\\
2, Rimskii-Korsakov Str., 244007 Sumy UKRAINE \\
E--mail: Alexander@olem.sumy.ua}

\maketitle

\begin{abstract}
The effective equation of motion that describes the
different monomer alternation along the heteropolymer chain
is proposed. On this
basis the supersymmetry field scheme is built up to analyze memory
and ergodicity breaking effects.

\end{abstract}

\pacs{PACS numbers: 61.41.+e, 05.40+j, 11.30.Pb}

Heteropolymer represents an ensemble of chains of
different type monomers (many--letter sequences), that can
experience the freezing and the microphase separation (MS)
(see [1] and references therein). Firstly, MS was studied
within random phase approximation [2] for the case of block copolymer
AB with arbitrary fraction $f$ of monomers of type A. At
$f\not= 0.5$ with temperature decrease the system
undergoes MS sequence of the first order, where period
$2\pi/k_0$ of space structure is of order of block length
and does not depend on temperature. By virtue of condition $k_0\not=
0$  the order parameter fluctuations give divergent
contribution to thermodynamic values, so that continuous
phase transition transforms to the weak first order transition [3].
Passing to random heteropolymers both MS and freezing transitions
remain, but here the space period is strongly dependent on temperature
[1, 4].  According to the field considerations [5, 6] the fluctuations
suppress both transitions in random copolymers,
but copolymer melts in disordered media suffer these transformations.

Similar to spin glass [7], the main tool of statistical mechanics of
random heteropolymers is the replica trick.
Besides, the transfer matrix method [8] and kinetic approach [9]
have been used. At the same time, it is well known
in theory of spin glass that within the framework of
Sherrington--Kirkpatrick model the replica method is identical
to the supersymmetry (SUSY) field approach
[10,11].  Being based on introducing Grassmann variables, the SUSY
field components have
explicit physical meaning, so that in contrast to the replica trick the
SUSY method arrives at physically interpreted results [12]. Up to
now, the SUSY method in polymer
theory has been proposed by Vilgis [13] and did not obtain further
development.

Usually, the SUSY method is based on the equation of motion
of Langevin type. But covalent bonds,
that ensure the formation of polymer chain, make such way
ineffective because the dynamic theory of polymers is
much more complicated than statistical mechanics of usual many--body
systems [14]. Therefore, one needs the effective motion equation
instead of dynamical one.  The main aim of this work is to obtain
such equation for random heteropolymers and to build up corresponding
statistical SUSY scheme.

Let us start with the directed polymer that represents Gaussian
chain with the probability $\Psi({\rm{\bf R}},N)$ to find the
end--point $N$ at the point ${\rm{\bf R}}$. As is known, the
function $\Psi({\rm{\bf R}},N)$ obeys the Schroedinger--type
equation with imaginary time $-{\rm i}N$ [14]
\begin{equation}
\partial\Psi/\partial N=
\left (D\partial^2/\partial{\rm{\bf R}}^2-U({\rm{\bf R}},N)\right
)\Psi, \label{1} \end{equation}
where number $N\gg 1$, $D$ is the effective diffusion coefficient, $U({\rm{\bf
R}},N)$ is the external field. In the limit where $N\to\infty$ the solution of
Eq.(1) can be written in the form of functional integral over
dependence ${\rm{\bf r}}(n)$ of chain coordinate on
number of internal monomers:  \begin{equation} \Psi({\rm{\bf
R}},N)=\int\exp(-S_{{\rm{\bf R}}N}\{{\rm{\bf
r}}(n)\}/2D)\delta{\rm{\bf r}}(n).  \label{2} \end{equation}
Here the action $S({\rm{\bf R}},N)\equiv S_{{\rm{\bf R}}
N}\{{\rm{\bf r}}(n)\}=\int_0^N L_0({\rm{\bf r}}(n)){\rm d}n$
with the fixed end--points ${\rm{\bf r}}(0)={\rm{\bf 0}}$
and ${\rm{\bf r}}(N)={\rm{\bf R}}$ is determined by Lagrangian
of the Euclidean field theory [10] \begin{equation} L_0={1\over
2}\left ({{\rm d}{\rm{\bf r}}(n)\over {\rm d}n}\right
)^2+2DU({\rm{\bf r}},n), \label{3} \end{equation}
where the effective
kinetic energy, for which the continuum limit ${\rm{\bf
r}}(n+1)-{\rm{\bf r}}(n)\to{\rm d}{\rm{\bf r}}(n)/{\rm d}n$ is assumed,
is caused by the covalent bond between monomers of polymer chain
[14].  Inserting Eq.(\ref{2}) to Eq.(1) gives the Jacobi--type
equation ${\partial S/\partial N}= D{\partial^2 S/\partial{\rm{\bf
R}}^2}-(1/2) \left ({\partial S/\partial {\rm{\bf R}}}\right
)^2+2DU$. After introducing the generalized momentum ${\rm{\bf
p}}\equiv\partial S/\partial{\rm{\bf R}}$ and the total derivative
${\rm d}{\rm{\bf p}}/{{\rm d}N}\equiv\partial{\rm{\bf p}}/\partial N+
({\rm{\bf p}}\partial/\partial{\rm{\bf R}}){\rm{\bf p}}$ the last
equation takes the linear form by Burgers: ${\rm d}{\rm{\bf p}}/{\rm
d}N=D\left (\partial^2{\rm{\bf p}}/\partial{\rm{\bf R}}^2+2\partial
U/\partial{\rm{\bf R}}\right )$. The above relations are well--known
in theories of directed polymers, stochastic growth and kinetic
roughening phenomena (see [15]).

Our basic observation [16] is that the
Schroedinger--type equation (\ref{1}) takes the form of the
Fokker--Planck equation \begin{equation} {\partial P\over\partial
N}=\left (D{\partial^2\over\partial{\rm{\bf
R}}^2}-{\partial\over\partial{\rm{\bf R}}}{\rm{\bf F}}\right )P
\label{4} \end{equation}
for the probability distribution
\begin{equation}
P({\rm{\bf R}},N)=\Psi({\rm{\bf R}},N)\exp\{-V({\rm{\bf R}})/2D\},
\label{5}
\end{equation}
where the dependence on ${\rm{\bf R}}$ is determined by the effective
potential $V\equiv -\int{\rm{\bf F}}{\rm d}{\rm{\bf R}}$.
Corresponding force ${\rm{\bf F}}$ is related to the
initial potential $U$ in Eq.(\ref{1}) as follows:
$U={\rm{\bf F}}^2/4D+(1/2){\partial{\rm{\bf
F}}/\partial{\rm{\bf R}}}$.
As is well known [17], Eq.(\ref{4}) determines the probability to
realize solution of corresponding Langevin--type equation
\begin{equation}
\partial{\rm{\bf
R}}/\partial N={\rm{\bf F}}({\rm{\bf R}},N)+{\rm{\bf \zeta}}(N)
\label{6}
\end{equation}
for stochastic function ${\rm{\bf R}}={\rm{\bf R}}(N)$. Here  the
stochastic force ${\rm{\bf\zeta}}$ is subjected to the
white--noise conditions $\langle\zeta(N)\rangle=0$,
$\langle\zeta(N)\zeta(N')\rangle=2D\delta (N-N')$,
where the angular brackets denote averaging with respect
to the distribution (\ref{5}).

In going from the case of directed polymer to the
 random heteropolymer AB, the
coordinate ${\rm{\bf R}}$ of the end--point $N$ becomes a stochastic
Ising variable $\theta(n)$, where $\theta(n)=1$ if $n$--th segment
is of type A and $\theta(n)=-1$
otherwise. Corresponding to the quenched disorder in fixed
sequence of different type segments, the law
$\theta(n)$ of monomer alternation along the chain
is described by the master equation analogous to
that determining the Glauber dynamics \cite{18}.
Relevant sequence correlator
$\overline{\sigma(n)\sigma(n')}$ for effective spin
$\sigma(n)\equiv\theta(n)-\overline{\theta(n)}$, being
deviation of microscopic value $\theta(n)$ from the average
$\overline{\theta(n)}$, takes the form \cite{8}
\begin{eqnarray}
&&\overline{\sigma(n)\sigma(n')}=C_2\exp(-|n-n'|/l),
\label{7}\\
&&C_2\equiv 4f(1-f),\qquad f\equiv(1/2)(1+\overline{\theta(n)}),
\nonumber
\end{eqnarray}

\noindent where overbar denotes the averaging over composition
sequence (quenched disorder), $l$ is the correlation length  and  $f$
is the fraction of type--A monomers.

Possessing the exponential correlator (\ref{7}), stochastic variable
$\sigma(n)$ is governed by the effective motion equation ${\rm
d}\sigma/{\rm d}n=-\sigma/l+s(n)$
with  the white noise:
$\overline{s(n)}=0$,
$\overline{s(n)s(n')}=2C_2l^{-1}\delta(n-n')$.
Relation between the microscopic value $\sigma(n)$ and stochastic
$\delta$--correlated variable $s(n)$ is given by the equality
\cite{17} $\sigma(n)=\int_0^n e^{-(n-m)/l}s(m){\rm d}m$.  Contrary to
the colored noise $\sigma(n)$, the white noise $s(n)$ has the
Gaussian distribution function
determining quenched disorder with the intensity
$C_2l^{-1}$. Then the locally averaging field
\begin{equation} \eta({\rm{\bf r}},n)\equiv (4C_2)^{-1/2}
\overline{\sigma(n)\delta({\rm{\bf r}}-{\rm{\bf r}}(n))}
\label{8} \end{equation}
represents the order parameter (hereafter the
monomer volume is assumed to be equal unity).

In addition to the terms of type included in in Eq.(\ref{6}),
the effective equation of motion for field (\ref{8}) must contain
the inhomogeneity
contribution $D\partial^2\eta/\partial{\rm{\bf
r}}^2$ that has
Fourier transform $-Dk^2\eta_{{\rm{\bf k}}}$.  As a result,
the equation of motion for the Fourier transformation $\eta_{{\rm{\bf k}}}(n)$
of field (\ref{8}) takes the Langevin form:
\begin{equation}
\partial \eta_{{\rm{\bf k}}}/\partial
n=-Dk^2 \eta_{{\rm{\bf k}}}- \partial
{\cal H}/\partial\eta^*_{{\rm{\bf k}}}+\zeta_{\rm{\bf k}}.
\label{9} \end{equation} Here, as above, the continuum approximation
is used for effective time $n\gg 1$, and the
definition of effective force $f_{{\rm{\bf k}}}=-{\partial
{\cal H}/\partial\eta^*_{{\rm{\bf k}}}}$ is taken
into account. The white noise $\zeta_{{\rm{\bf k}}}=\zeta_{{\rm{\bf
k}}}(n)$ is defined by the conditions
\begin{equation}
\langle\zeta_{{\rm{\bf k}}}\rangle=0,\qquad \langle\zeta_{{\rm{\bf
k}}}^*(n)\zeta_{{\rm{\bf k'}}}(n')\rangle=2\delta_{{\rm{\bf
k}}{\rm{\bf k'}}}\delta(n-n'), \label{10} \end{equation}
where the
angular brackets denote averaging over chain conformations (thermal
disorder).  Taking into account the effect of
fluctuations, the effective Hamiltonian reads
\cite{1,5,8} \begin{eqnarray}
&&{\cal H}=\sum_{{\rm{\bf
k}}}r_{{\rm{\bf k}}}|\eta_{{\rm{\bf k}}}|^{2}-{1\over
2}\sum_{{\rm{\bf k}}{\rm{\bf k'}}}w_{{\rm{\bf k}} {\rm{\bf
k'}}}|\eta_{{\rm{\bf k}}}|^2|\eta_{{\rm{\bf k'}}}|^2+\int v({\rm{\bf
r}}){\rm d}{\rm{\bf r}}; \label{11}\\ &&r_{{\rm{\bf k}}}\equiv
r+2D(k-k_0)^2, \qquad k_0^{-1}\equiv 2(\pi D)^{1/2}l(2r)^{1/4},
\nonumber\\
&&r=\tau+(3/4\pi)l^{-2}(2r)^{-1/2},\qquad \tau\equiv
l^{-1}-C_2\chi;
\nonumber\\
&&w_{{\rm{\bf k}}{\rm{\bf k'}}}\equiv
4\sigma^2 l^{-2}(ND)^{-1}({\rm{\bf k}}^2+{\rm{\bf
k'}}^2)^{-1};\nonumber\\
&&v=-(\mu/3!)\eta^3+(\lambda/4!)\eta^4,\nonumber\\
&&\mu\equiv 12C_3C_2^{-1/2}l^{-1},\qquad
\lambda\equiv24(1+5C_3^2/C_2)l^{-1}, \nonumber\\ &&C_2\equiv
4f(1-f),\qquad C_3\equiv|1-2f|.  \nonumber \end{eqnarray}
Here $\sigma$, $\chi$ are the interreplica overlapping and
the Flory parameters, correspondingly.

The following application of the SUSY scheme is straightforward
\cite{10}. One has to introduce the generating functional
\begin{eqnarray}
&&Z\{\eta_{\rm{\bf
k}}\}=\left\langle\delta\left ({\partial \eta_{\rm{\bf
k}}\over\partial n}+{\delta {\cal H}\over \delta \eta_{\rm{\bf
k}}^*}-\zeta_{\rm{\bf k}}\right )\det\left|{\delta\zeta_{\rm{\bf
k}}\over \delta \eta_{\rm{\bf k}}}\right|\right\rangle, \label{12}\\
&&{\delta {\cal H}/ \delta \eta_{\rm{\bf k}}^*}\equiv{\partial
{\cal H}/ \partial \eta_{\rm{\bf
k}}^*}+2D(k-k_0)^2\eta_{\rm{\bf k}},\nonumber
\end{eqnarray}
being the average over noise $\zeta_{{\rm{\bf k}}}(n)$, where
$\delta$--function accounts for the motion equation (9), the
determinant is Jacobian of transformation from $\zeta_{{\rm{\bf k}}}$ to
$\eta_{{\rm{\bf k}}}$. Then the functional Laplace representation is
used for $\delta$--function, that gives a ghost field
$\varphi_{{\rm{\bf k}}}(n)$.
One needs to use Grassmann
conjugated fields $\psi_{{\rm{\bf k}}}(n)$, $\bar\psi_{{\rm{\bf
k}}}(n)$ \cite{10} in order to write the
determinant in exponential form. Then, assuming that
conformation averaging in Eq.(\ref{12}) is Gaussian
with variance 1 (see Eqs.(\ref{10})), the standard form is derived
(cf. Eqs.(2,3))
\begin{eqnarray} && Z\{\eta\}=\int
P\{\eta,\varphi;\psi,\bar{\psi}\}\delta\varphi\delta^2\psi;
\qquad \delta^2\psi\equiv\delta\psi\delta\bar\psi,
\label{13}\\ &&
P\{\eta,\varphi;\psi,\bar{\psi}\}=\exp(-S\{\eta,\varphi;\psi,\bar{\psi}\}),\qquad S=\int_0^N L{\rm d}n,
\nonumber\\
&& L=\int[(\varphi\dot
\eta+\bar{\psi}\dot\psi-\varphi^2/2)+({\cal
H}'\{\eta\}\varphi+\bar{\psi}{\cal H}''\{\eta\}\psi)]{\rm d}{\rm{\bf
r}}.  \nonumber \end{eqnarray} Here, the point denotes the derivative
with respect to
"time" $n$, the prime denotes the functional derivative with respect
to the field (8).

The last expression in Eqs.(\ref{13}) takes the simplest form
\begin{eqnarray} && L={1\over 2}\int \Lambda (\Phi){\rm
d}^2\vartheta,\qquad \Lambda\equiv\sum_{\rm{\bf k}}\Phi_{\rm{\bf
k}}^*\bar{{\cal D}}{\cal D}\Phi_{\rm{\bf k}}+{\cal H}\{\Phi_{\rm{\bf
k}}\}, \label{14}\\&&{\cal
D}\equiv{\partial/\partial\bar{\vartheta}}+\vartheta{\partial/\partial n},
\qquad \bar{{\cal D}}\equiv{\partial/\partial
\vartheta}+\bar{\vartheta}{\partial/\partial n} \nonumber
\end{eqnarray}
within the SUSY field representation
\begin{equation}
\Phi=\eta+\bar{\psi}\vartheta+\bar{\vartheta}\psi+\bar{\vartheta}\vartheta\phi,
\label{15}
\end{equation}
where Grassmann coordinates $\vartheta,\bar{\vartheta}$ obey the
same relations as fields $\psi,\bar{\psi}$. Here the new
field ${\phi}_{\rm{\bf k}}\equiv\dot\eta_{\rm{\bf
k}}-\varphi_{\rm{\bf k}}$ is introduced,
the functional ${\cal H}\{\Phi\}$ has the same form as the effective
Hamiltonian (\ref{11}), where the order parameter $\eta_{\rm{\bf k}}$ is
replaced by the SUSY field $\Phi_{\rm{\bf k}}$, Eq.(\ref{15}), and the term
$(\delta {\cal H}/\delta\eta_k^*) \dot{\eta_{\rm{\bf k}}}+c.c.$
is the total derivative ${\rm d}{\cal H}\{\eta_{\rm{\bf
k}}(n)\}/{\rm d}n$ that can be omitted.  According to
\cite{12}, the physical meaning of the SUSY field components is as
follows: $\phi\equiv -\delta {\cal H}/\delta\eta^*$ is the field
being conjugated to the order parameter $\eta$,
$\varphi\equiv\dot\eta-\phi$ is the most probable value of
fluctuation of the conjugate field, and the combination $\psi\bar{\psi}$
gives the density of sharp interphases. So, the using of the whole
4--component SUSY field (\ref{15}) corresponds to the strong
segregation limit. In what follows we concentrate on the simple case of
the weak segregation limit where $\psi\bar{\psi}\equiv 0$. Then, the
SUSY field (\ref{15}) is reduced to the 2--component form
\begin{equation} \Phi=\eta+{\theta}\varphi,
\label{16} \end{equation} where self--conjugated nilpotent
variable ${\theta}\equiv\bar{\vartheta}\vartheta$ is introduced.
As a result,  the Lagrangian (\ref{14}) takes the form
\begin{eqnarray}
&& L={1\over 2}\int\Lambda(\Phi){\rm d}{\theta}, \qquad \Lambda\equiv
\sum_{\rm{\bf k}}\Phi_{\rm{\bf k}}^*D\Phi_{\rm{\bf k}}+
{\cal H}\{\Phi_{\rm{\bf k}}\}; \label{17}\\
&& D=-{\partial/\partial\theta}+\left
(1-2\theta{\partial/\partial\theta}\right )
{\partial/\partial n}. \nonumber \end{eqnarray}
Here the
fluctuation field $\varphi$ is used instead of conjugate field
$\phi$ and the lowest power of derivative $\partial/\partial n$ is
retained. To correspond to Lagrangian (\ref{17}) the motion equation
for SUSY field (\ref{16}) reads
\begin{equation} D\Phi_{\rm{\bf k}}=-\delta
{\cal H}/\delta\Phi_{\rm{\bf k}}^*. \label{18} \end{equation}

Let us introduce now the SUSY correlator
\begin{equation}
C_{\rm{\bf k}}(n,{\theta};n',\theta ')\equiv
\left\langle\Phi_{\rm{\bf k}}^*(n,\theta)\Phi_{\rm{\bf
k}}(n',\theta ')\right\rangle\vartheta(n-n'), \label{19}
\end{equation}
where $\vartheta(n-n')=1$ if $n>n'$ and $\vartheta(n-n')=0$
otherwise. Multiplying Eq.(\ref{18}) by $\Phi_{\rm{\bf
k}}^*$ and averaging the result, one gets within zeros approximation
($w=v=0$ in Eqs.(\ref{11})) \cite{19}
\begin{equation}
C_{\nu{\rm{\bf k}}}^{(0)}(\theta,\theta ')={1+(r_{\rm{\bf
k}}-{\rm i}\nu)\theta+(r_{\rm{\bf k}}+{\rm i}\nu)\theta
'\over r_{\rm{\bf k}}^2+\nu^2}.  \label{20} \end{equation}
Here,
conventional frequency $\nu$ denotes Fourier transformation over
"time" being the monomer number $n$.
The most important feature of expression
(\ref{20}) is the characteric structure with respect to combination
of the nilpotent variables $\theta$, $\theta '$, that is
inherent not only zeros approximation but arbitrary
SUSY correlator. To this end it is convenient to introduce
basis supervectors
\begin{equation} A(\theta,\theta ')=\theta, \qquad
B(\theta,\theta ')=\theta ',\qquad T(\theta,\theta ')=1, \label{21}
\end{equation}
containing functional product $X (\theta,\theta ') =
\int  Y(\theta,\theta '')$ $Z(\theta '',\theta) {\rm
d}\theta ''$ for any SUSY vectors ${\bf X}$, ${\bf Y}$,
${\bf Z}$. It is easy to see, that basis SUSY vectors (\ref{21})
obey the following multiplication rules: ${\bf A}^2={\bf A}$,
${\bf B}^2={\bf B}$, ${\bf B}{\bf T}={\bf T}$, ${\bf T}{\bf
A}={\bf T}$, other products are zero.
Since the set of
vectors ${\bf A}$, ${\bf B}$, ${\bf T}$ is closed, it
is convenient to expand any SUSY correlator over this basis:
\begin{equation}
{\bf C}=G_+ {\bf A} + G_- {\bf B} + S{\bf T}. \label{22}
\end{equation}
Hereafter the subscripts ${\rm{\bf k}}$, $\nu$ are suppressed for
brevity.  Using Eqs.(\ref{16}), (\ref{19}), one gets for coefficients
of expansion (\ref{22}): $G_+=\langle\varphi^*\eta\rangle$,
$G_-=\langle\eta^*\varphi\rangle$,
 $S=\langle|\eta|^2\rangle$.
So, $G_{\pm}$ represent advanced and
retarded Green functions and $S$ is the structure factor. In
accordance with Eqs.(\ref{20}), (\ref{21}), (\ref{22}), within
zeroes approximation these functions are
\begin{equation}
G_\pm^{(0)}=(r\pm{\rm i}\nu)^{-1},
~~~~S^{(0)}=G_+^{(0)}G_-^{(0)}=(r^2+\nu^2)^{-1}. \label{23}
\end{equation}

The Dyson equation for the SUSY correlator (\ref{19})
reads \cite{12}
\begin{equation}
{\bf C}^{-1}=({\bf C}^{(0)})^{-1}-w{\bf C}-{\bf \Sigma},
\label{24}
\end{equation}
where $w=2\cdot6^{1/3}\pi^{-4/3} D^{-1}(\sigma/l)^2$ is the typical
value of the sum $\sum_{{\bf k}'} w_{{\bf k}{\bf k}'}$ in Hamiltonian
(\ref{11}), ${\bf \Sigma}$ is the SUSY self--energy operator.
Using Eqs.(\ref{23}) and
expansions  (\ref{22}),
\begin{equation} {\bf
\Sigma}=\Sigma_+{\bf A}+\Sigma_-{\bf B}+\Sigma{\bf T}, \label{25}
\end{equation}
the Dyson equations for the components $G_\pm$, $S$
become

\begin{equation}
G_\pm^{-1}+wG_\pm=(r\pm{\rm i}\nu)-\Sigma_\pm,
\eqnum{26a}\label{26a}
\end{equation}

\begin{equation}
S=(1+2\pi C_2l^{-1}\delta(\nu)+\Sigma)G_+G_-(1-wG_+G_-)^{-1}.\label{26b}
\eqnum{26b}
\end{equation}
Here $\delta$--term appears to take into account
the contribution to Lagrangian due to quenched
disorder. Coefficients of SUSY expansion (\ref{25})
are \cite{19}

\begin{equation}
\Sigma_\pm(n)=
\left(\mu^2+{\lambda^2\over 2}S(n)\right)S(n)G_\pm(n),\eqnum{27a}
\label{27a} \end{equation}

\begin{equation}
\Sigma(n)=\left(\mu^2+{\lambda^2\over
6}S(n)\right)S^2(n),\eqnum{27b} \label{27b}
\end{equation}
where the ${\rm{\bf r}}$--representation is used for
macroscopically homogeneous system.

Let us introduce now the memory parameter $q\equiv\langle
\eta(n=N)\eta(n=0)\rangle$ and the non--ergodicity parameter
$\Delta\equiv g_0-g$ that is the difference between isothermal
susceptibility $g_0\equiv G_-(\nu=0)$ and thermodynamic value
$g\equiv G_-(\nu\to 0)$. Being coefficients in SUSY
expansion (\ref{22}), the main correlators of non--ergodic system with
memory acquire the elongated form:

\setcounter{equation}{27}
\begin{equation}
G_\pm(\nu)=\Delta+G_{\pm 0}(\nu),\qquad S(n)=q+S_0(n), \label{28}
\end{equation}
where the index $0$ denotes the components corresponding to ergodic
system without memory. Inserting Eqs.(\ref{28}) to Eqs.(27)
provides the self--energy components $\Sigma_\pm(n)$, $\Sigma(n)$
as functions of parameters $\Delta$, $q$ within
the "time"--representation.
Since the Dyson equations (26) require "frequency" Fourier
transforms, it is convenient to use the fluctuation--dissipation
theorem $S_0(n\to 0)=G_{\pm 0}(\nu\to 0)\equiv g$,
$\Sigma_{\pm 0}(\nu\to 0)=\Sigma_0(n\to 0)$. As a result, the
Dyson equation (\ref{26b}) gives in the $\nu$--representation
\begin{equation}
q\left[1-wg_0^2-{1\over 2}\left(\mu^2+{\lambda^2\over
3}q\right)qg_0^2\right]=C_2l^{-1}g_0^2,
\label{29}
\end{equation}

\begin{equation}
S_0={{(1+\Sigma_0)G_+G_-}\over{1-[w+(\mu^2+\lambda^2q/2)q]G_+G_-}}.
\eqnum{29a} \label{29a}
\end{equation}
The first of these equations corresponds to the $\delta$--term
due to memory effects, the second one corresponds to "frequency"
$\nu\not=0$. At $\nu\to 0$ the characteristic product is
$G_+G_-\to g^2$, so that the pole of the structure factor (29a)

\setcounter{equation}{29}
\begin{equation}
w+\left(\mu^2+{\lambda^2\over 2}q\right)q=g_0^{-2}
\label{30} \end{equation}
determines the point of ergodicity breaking. Analogously,
Eq.(\ref{26a}) gives the equation for the susceptibility
$g\equiv G_-(\nu\to 0)$
\begin{equation}
wg^2+{\mu^2\over 2}g\left[(g+q)^2-q^2\right]+{\lambda^2\over
6}g\left[(g+q)^3-q^3\right]=rg-1.
\label{31} \end{equation}
The system of equations (29)--(\ref{31}) completely describes
thermodynamic behaviour of random heteropolymer by analogy with
the spin glass theory \cite{7}. So, Eqs.(29), (\ref{31}) play a role
of Sherrington--Kirkpatrick equations, and Eq.(\ref{30}) defines the
point of de Almeida--Thouless instability.

According to Eqs.(29), (\ref{30}), the memory parameter is given
by the cubic equation
\begin{equation}
(\mu^2/2+\lambda^2q/3)q^2=C_2l^{-1}.
\label{32} \end{equation}
Taking into account definitions (\ref{11}), it is seen
that for copolymers close to symmetric composition $f=0.5$ ($C_3\ll
C_2$) the first term in brackets of Eq.(\ref{32}) is negligible, and
the dependence $q\propto l^{1/3}$ takes place. In the opposite case of
dilute copolymer, where $f\ll 1$ ($C_2\ll C_3$), we have the
less value $q\propto fl^{1/2}$. Naturally, the memory parameter
diminishes always with correlation length decreasing.

The behaviour of isothermal, $g_0$, and thermodynamic, $g$,
susceptibilities is described by the equations (\ref{30}) and (\ref{31}),
respectively. The
magnitude $g_0$ depends only on the quenched disorder $l$, whereas the
latter, $g$, is defined by both values $l$, $\chi$ of quenched and
thermal disorders (see Eqs.(\ref{11}) for dependence $r(\chi)$). The
point of ergodicity breaking, $r_0$, is determined by equation
\begin{equation} 2-rg_0+{\mu^2\over 2}g_0^3+{\lambda^2\over
6}g_0^3(g_0+3q)=0, \label{33} \end{equation} that follows from
Eqs.(\ref{30}), (\ref{31}) at $g=g_0$. At the point of MS, $r_c$, one
has ${\rm d}g/{\rm d}r=-\infty$, and Eq.(\ref{31}) gives the condition
\begin{equation}
w+\mu^2(g+q)+{\lambda^2\over 2}(g+q)^2=g^{-2}.
\label{34}
\end{equation}
The form of phase diagram given by Eqs.(\ref{33}), (\ref{34}) is
shown in Fig.1.

\begin{figure}
\caption{
The phase diagram for random copolymer with $\sigma=0$
(a) and $\sigma=5$ (b) ( $l=1$, $D=1$; solid and broken curves are
the MS and ergodicity breaking lines correspondingly)
}
\label{fig1}
\end{figure}


\begin{thebibliography}{00}

\bibitem{1} C.D. Sfatos, E.I. Shakhnovich, {\it Phys. Rep.} {\bf
288}, 77 (1997).

\bibitem{2} L. Leibler, {\it Macromolecules}  {\bf 13}, 1602 (1980).

\bibitem{3} G.H. Fredrickson, E. Helfand, {\it J. Chem. Phys.} {\bf
87}, 697 (1987).

\bibitem{4} V.S. Pande, A.Yu. Grosberg, T. Tanaka, {\it Phys. Rev.}
{\bf E51}, 3381 (1995).

\bibitem{5} A.V. Dobrynin, I.Ya. Erukhimovich, {\it J. Phys. I
France} {\bf 5}, 365 (1995).

\bibitem{6} S. Stepanov, A.V. Dobrynin, T.A. Vilgis, K. Binder, {\it
J.  Phys. I France} {\bf 6}, 837 (1996).

\bibitem{7} M. Mezard, G. Parisi, M.A. Virasoro, {\it Spin Glass
Theory and Beyond} (World Scientific, Singapore, 1987).

\bibitem{8} G.H. Fredrickson, S.T. Milner, L. Leibler, {\it
Macromolecules} {\bf 25}, 6341 (1992).

\bibitem{9} E.G. Timoshenko, Yu.A. Kuznetsov, K.A. Dawson, {\it Phys.
Rev.} {\bf E54}, 4071 (1996).

\bibitem{10} J. Zinn--Justin, {\it Quantum Field Theory and
Critical Phenomena} (Clarendon Press, Oxford, 1993).

\bibitem{11} J. Kurchan, {\it J. Phys. I France} {\bf 2} 1333 (1992).

\bibitem{12} A.I. Olemskoi, I.V. Koplyk, {\it Physics--Uspekhi} {\bf
38}, 1061 (1995).

\bibitem{13} T.A. Vilgis, {\it J. Phys.} A{\bf 24}, 5321 (1991).

\bibitem{14} M. Doi, S.F. Edwards, {\it The Theory of Polymer
Dynamics} (Clarendon Press, Oxford, 1986).

\bibitem{15} T. Halpin--Healy, Y.--C. Zhang, {\it Phys. Rep.} {\bf
254}, 215 (1995).

\bibitem{16} A.I. Olemskoi, {\it Uspekhi Fizicheskikh Nauk}
{\bf 168}, 287 (1998).

\bibitem{17} H. Risken, {\it The Fokker--Planck Equation} (Springer,
Berlin--Heidelberg, 1989).

\bibitem{18} R.J. Glauber, {\it J. Math. Phys.} {\bf 4}, 294 (1963).

\bibitem{19} A.I. Olemskoi, V.A. Brazhnyi, {\it to be published}.

\end{thebibliography}
\end{document}